\documentclass[twocolumn,floats,pra]{revtex4}

\newcommand {\be}{\begin{equation}}
\newcommand {\ee} {\end{equation}}
\newcommand {\bea}{\begin{eqnarray}}
\newcommand {\eea} {\end{eqnarray}}

\begin{document}


\title{Conformal invariance of chiral edge theories}
\author{N. Read}
\affiliation{Department of Physics, Yale University, P.O. Box
208120, New Haven, CT 06520-8120, USA}
\date{March 6, 2009}

\begin{abstract}
The low-energy effective quantum field theory of the edge
excitations of a fully-gapped bulk topological phase corresponding
to a local Hamiltonian must be local and unitary. Here
it is shown that whenever all the edge excitations propagate in
the same direction with the same velocity, it is a conformal field
theory. In particular, this is the case in the quantum Hall effect
for model ``special Hamiltonians'', for which the ground state,
quasihole, and edge excitations can be found exactly as
zero-energy eigenstates, provided the spectrum in the interior of
the system is fully gapped. In addition, other conserved
quantities in the bulk, such as particle number and spin, lead to
affine Lie algebra symmetries in the edge theory. Applying the arguments to some trial wavefunctions related to non-unitary conformal field theories, it is argued that the Gaffnian state and an infinite number of others cannot describe a gapped topological phase because the numbers of edge excitations do not match any unitary conformal field theory.
\end{abstract}

\pacs{PACS numbers: } 

\maketitle


\section{Introduction}

There has been renewed interest recently in fractional quantum
Hall systems, especially non-Abelian ones \cite{mr}, in connection
with topological quantum computing. Such systems are examples of
topological phases in two space dimensions, that is systems with a
gap for all bulk excitations (i.e.\ those far from any boundary),
and no apparent broken symmetries or conventional (i.e.\ local)
order parameter. Topological phases can be distinguished by the
quantum numbers and statistics (defined by adiabatic exchange) of
their point-like excitations \cite{mr}. Topological phases in
general may possess gapless excitations at an edge, that is the
boundary with vacuum outside, or also at an interface between
regions of the system in distinct topological phases. When there
are gapless edge excitations, they can be described by some
effective low-energy theory. This theory has generally been
expected to be a conformal field theory (CFT). However detailed
arguments for this appear to be lacking in the condensed matter
situation, in which for example Lorentz invariance cannot be taken
for granted.

In this note we discuss conformal invariance properties of an
effective field theory of the edge, with particular reference to
quantum Hall (QH) systems. We show that when the bulk (interior) of
the system is fully gapped and all edge excitations propagate in
the same direction with the same velocity, the edge theory is a
local unitary conformal field theory. These conditions are
sometimes met by the ``special Hamiltonians'', which exist for
some of the bulk topological QH phases; these have zero-energy
eigenfunctions that can be found explicitly, and which are given
by correlation functions in some CFT which is also the natural
candidate for the corresponding effective edge theory \cite{mr}.
At the end we apply these remarks to disqualify some recently-proposed trial wavefunctions from describing a topological phase.

\section{Special Hamiltonians}

First we address the special Hamiltonians just mentioned. Many
examples of these are now known, see for instance
\cite{hald,hr,milr,rr1,rr2} (in general we can allow the particles
to have internal ``spin'' quantum numbers, and corresponding
symmetries of the Hamiltonian may act on these indices). In each
case, there is a short-range interaction for particles in the
lowest Landau level (LLL). For a QH system in the infinite plane,
the LLL states for a single particle in the symmetric gauge are
spanned by the functions $z^m e^{-|z|^2/4}$ with $m=0$, $1$,
\ldots (we have set the magnetic length to 1). As the
single-particle states in the LLL are degenerate, the
many-particle Hamiltonian $H_0$ can be taken to be solely the
interaction, acting in the space of many-particle states in the
LLL. Then, for suitable $H_0$, the full space of zero-energy
eigenfunctions in the infinite plane can be obtained (possibly
only after some effort). The simplest of all
examples is%
\be%
H_0=V_0\sum_{i<j}\delta^2(z_i-z_j),\label{haldham}\ee%
projected to the LLL \cite{hald}, and acting in the space of
symmetrized LLL wavefunctions, describing bosons; $V_0>0$ is a
constant. In some cases the special Hamiltonian contains many-body
(say, up to some fixed $k$ independent of the particle number $N$)
local interactions. This Hamiltonian is not intended to be
directly physical, but serves as a paradigm and an existence proof
for a particular phase of matter. The angular momentum $M$ is
defined in two dimensions for the LLL as $M=\sum_iz_i
\partial/\partial z_i$ (where the derivatives act only on the
analytic part of the wavefunction), and always commutes with the
Hamiltonian. As in the case of the Hamiltonian, it can be viewed
as an integral of a local operator in the two-dimensional space. We
will consider only the cases in which there is a unique
zero-energy eigenstate of lowest angular momentum $M=M_0$, which
corresponds to the ground state of the phase in question, in the
form of a circular drop (disk) of radius $R$ as in Laughlin's
original work \cite{laugh}. (In the example of $H_0$ above, this
ground state is the Laughlin state of exponent $2$ for bosons at
filling factor $\nu=1/2$.) ``Low-lying'' excitations above this
ground state are defined as zero-energy eigenstates with a small
increase in $M$; more formally, that is $\Delta M=M-M_0$ of order
$1$ as the particle number $N\to\infty$. For these it is useful to
add to the special Hamiltonian $H_0$ the term $\omega M$ for
$\omega>0$. Then as $M$ commutes with $H=H_0+\omega M$, the zero
energy eigenstates are unchanged, and span subspaces, with one
subspace for each value of $M$. Then the energy of the eigenstates
increases with $M$, but is the same for each $M$.

We will introduce one further crucial assumption about the spectrum
of $H_0$: in the bulk of the system (as $N\to\infty$), there is a
gap that converges to a nonzero constant in the thermodynamic
limit $N\to\infty$. This is not the case for some known special
Hamiltonians, but based on numerical work does appear to be the
case for others. When this requirement holds, one can argue that
the low-lying zero-energy eigenstates of $H_0$ represent
excitations located at the edge of the circular drop describing
the system. (Note that other states associated with particles far
outside the drop will have energies approaching zero at high $M$.)
For the Hamiltonian $H=H_0+\omega M$, these possess energies above
that of the ground state, but only by amounts of order
$\omega\Delta M$, so $\omega>0$ is an angular velocity. Then the
velocity of propagation of the edge excitations is $v=\omega R$.
There are also quasihole excitations of the system, which are
zero-energy eigenstates with angular momentum $\Delta M$ of order
$1$ as $N\to\infty$.

\section{Effective field theory}

Now let us consider this system from the point of view of an
effective quantum field theory for the edge. This field theory has
one space and one time dimension; the space dimension is periodic,
of course. For the limit of large systems, the circumference of
the edge becomes arbitrarily large, and we can ignore this
periodic boundary condition, and consider properties of the theory
locally on the edge. The quantum field theory is unitary. A
unitary field theory may be defined as one that has both a
positive-definite inner product on its space of states (making
this space a true Hilbert space), and a Hamiltonian that is
self-adjoint with respect to this inner product. Both properties
are required in all conventional quantum mechanical systems, and
for the effective theory of our edge both continue to hold when we
integrate out other higher energy states (non-zero-energy states
of $H_0$) to derive the effective Hamiltonian. Due to the locality
of the underlying Hamiltonian $H$, and the gap in the bulk
excitation spectrum, the effective Hamiltonian will be an integral
along the edge of a local operator $T/(2\pi)$ (the factor $2\pi$
here and below is conventional in CFT); thus the effective theory
is also local. Local observables of the original system will also
remain local when the corresponding operators in the effective
edge theory are derived. For a local quantum field theory, we may
refine our definition of unitarity by replacing it with the more
stringent condition that the Hamiltonian density, or better the
stress-energy-momentum tensor, is self-adjoint. The
stress-energy-momentum tensor $T_{\mu\nu}$ of a field theory,
which will be called the stress tensor for short, is the Noether
current operator associated with translational invariance in space
and time; here $\mu$, $\nu=x$, $t$, where $x$ is a coordinate
along the edge. In general, the Hamiltonian density is the
time-time component $T_{00}/(2\pi)$ of the stress tensor.

For the edge theory, all excitations are chiral, that is they
propagate along the edge, with the common velocity $v$ at
sufficiently low energies. Then for any local operator $\cal O$ in
the effective field theory, we have the equation of motion
$\partial_t{\cal O}=-v\partial_x{\cal O}$, and so $\cal O$ is a
function only of $x-vt$. Let us redefine $z=x-vt$, and choose
units so that $v=1$.

For the stress tensor, one conventionally defines the light cone
components with respect to the coordinates $z=x-t$,
$\overline{z}=x+t$ \cite{bpz,cft}. Here $z$ and $\overline z$ are
independent real numbers. [It is sometimes useful to analytically
continue by allowing $x$ and $t$ to be complex, writing
$t=-i\tau$; conventional imaginary time (or Euclidean spacetime)
is obtained by requiring $x$, $\tau$ to be real, so that $z$ is
complex and $\overline z$ is the complex conjugate of $z$.] The
Hamiltonian density becomes $T_{00}/(2\pi)=[T_{zz}+T_{\overline
z\overline z}]/(2\pi)$, and the momentum density is
$T_{11}/(2\pi)=[T_{zz}-T_{\overline z\overline z}]/(2\pi)$. The
continuity equation obeyed by $T_{\mu\nu}$ becomes
$\overline\partial T_{zz}+\partial T_{\overline z z}=0$, where
$\partial=\frac{1}{2}(\partial_x-\partial_t)$,
$\overline\partial=\frac{1}{2}(\partial_x+\partial_t)$; this
corresponds to conservation of energy and momentum. In addition,
invariance under rotations of space (in space dimension larger
than one) and under Lorentz boosts requires that $T_{\mu\nu}$ be
symmetric, which becomes simply $T_{z\overline z}=T_{\overline z
z}$ in light-cone variables in two dimensions.

In the present case, the Hamiltonian density $T/(2\pi)$
obeys%
\be%
\overline\partial T=0,\ee%
and can be identified as $T=T_{zz}$, while $T_{\overline
z\overline z}=T_{z\overline z}=T_{\overline z z}=0$. This has
striking consequences. First, in addition to conservation of
energy and linear momentum (the latter corresponds to angular
momentum for the circular edge), we also have Lorentz/scale
invariance. In the light cone coordinates, a Lorentz boost acts as
$z\to e^\theta z$, $\overline z\to e^{-\theta}\overline z$, where
$\theta$ is real, while for a scale transformation, $z$ and
$\overline z$ are both scaled by the same factor. If we consider
the transformation on $z$ only, then these transformations cannot
be distinguished. The generator for either is the integral
(divided by $2\pi$) of the chiral dilatation current $j_{Dz}=T z$,
which is divergenceless, $\overline\partial j_{Dz}=0$, so $\int
dx\, j_{Dz}$ is conserved. If we restore the periodic boundary,
the closest analogue of $j_{Dz}$ is $T R\sin (z/R)$, for the drop
of radius $R$. The Lorentz boost produces a Lorentz contraction of
the circumference $2\pi R$ of the drop; there is a natural rest
frame in which the circumference is maximal (but one must examine
both $z$ and $\overline z$ to see this geometry).

For a general discussion of the relation between scale and
conformal invariance in a Lorentz-invariant field theory, see
Ref.\ \cite{pol}. For our purposes a useful sufficient condition
for both is the existence of a symmetric stress tensor with
vanishing trace, $T_\mu^\mu=0$ (where the summation convention is
in effect, and indices are raised using the canonical inverse
metric preserved by Lorentz transformations), which in light-cone
variables means $T_{z\overline z}=0$. (We say ``existence'' because
the stress tensor is not uniquely defined, and one can consider
alternatives that differ by certain terms but all give the same
Hamiltonian \cite{pol}; we can use any one such that
$\overline\partial T_{zz}=0$.)

Conformal transformations in two dimensions form an
infinite-dimensional algebra of symmetries (it includes the
Lorentz and scale transformations) \cite{bpz,cft}. An
infinitesimal conformal mapping is described by a vector field
with components $v^z(z)$, $\overline{v}^{\overline z}(\overline
z)$, where $\overline
\partial v^z=\partial \overline v^{\overline z}=0$, so $v^z(z)$
[$\overline{v}^{\overline z}(\overline z)$] can be viewed as a
function of $z$ ($\overline z$) only; the transformation is $z\to
z+\varepsilon v^z(z)$ (and similarly for $\overline z$) for
$\varepsilon$ small, and $v^z=1$, $z$ yield the preceding special
cases. (Each vector field should be analytic in some open set in
the complex $z$ plane.) Again, the existence of a stress tensor
with vanishing trace is a sufficient condition for invariance
\cite{pol}. For our case, a divergenceless current corresponding
to the transformation by $v$ is $j_{vz}(z)=v^z(z)T_{zz}$, which
again obeys $\overline\partial j_{vz}=0$. Indeed, perhaps it
should have been obvious from the beginning that the purely chiral
dynamics would lead in the thermodynamic limit to an infinite
number of conserved quantities; we are presently studying only
those that generate space-time transformations.

We now have an algebra of operators $Q_v=\int dx\,j_{vz}/(2\pi)$
acting in the Hilbert space of the edge theory that forms a
representation of the algebra of infinitesimal conformal mappings.
The relations in this algebra are determined by the commutation
relations of $T$, for which the most
general form is (see e.g.\ Ref.\ \cite{aff85})%
\bea%
[T(z),T(z')]/(2\pi i)&=&\delta(z-z')\partial
T(z')-2(\partial\delta(z-z'))T(z')
\nonumber\\
&&\mbox{}+(c/12)\partial^3\delta(z-z').\eea %
The first two terms follow from the facts that $T$ generates
translations in $z$ and has scaling dimension (conformal weight)
$2$. For two vector fields $v^z$, $w^z$, the commutator of $Q$'s is %
\be%
[Q_{v},Q_{w}]=-iQ_{[v,w]}+(ic/24\pi)\int dz\,v^z\partial^3w^z,\ee%
where $[v,w]^z=v^z\partial w^z-w^z\partial v^z$, the Lie bracket
of $v$ and $w$, is another vector field. The terms other than that
containing $c$ are the relations defining the Lie algebra of
conformal mappings. The additional term containing only the scalar
$c$ times the identity operator is allowed for the usual reason
that a given symmetry algebra may be represented projectively in
quantum mechanics; in the present case, this term is unique modulo
gauge transformations of the quantum states, under which $c$ is
gauge invariant, so the term cannot be gauged away unless $c=0$
\cite{gf}. Returning to the system with periodic boundary
condition, we form the Fourier modes (at $t=0$) $L_n=\oint dx\,
e^{-inx/R}\, TR/(2\pi)+\delta_{n,0}c/24$. The corresponding
algebra of these
modes is the Virasoro algebra, with relations \cite{bpz,cft}:%
\be%
[L_n,L_m]=(n-m)L_{n+m}+\frac{c}{12}n(n^2-1)\delta_{n+m,0}.\ee%
The central charge $c$ is a constant within a given edge
theory. Self-adjointness of $T$ implies that the modes satisfy $L_n^\dagger=L_{-n}$. In a unitary theory, $c$ must be positive; no unitary CFTs exist for $c\leq 0$, except the trivial theory at $c=0$, in which the only state is the vacuum, and the stress tensor is zero. Thus we have established that the effective theory of the edge is a
local unitary chiral CFT (in particular, it is relativistic, i.e.\
Lorentz invariant) \cite{cft}.

To complete the discussion of conformal invariance, we should show
that there is a ``vacuum'' state $|0\rangle$ in the edge theory
that is invariant under the M\"obius group, the group generated by
the Virasoro generators $L_0$, $L_{-1}$, $L_{+1}$; then
correlation functions in this vacuum possess M\"obius invariance
properties \cite{bpz,cft}. The vacuum is also annihilated by the
Fourier modes $L_n$ for all the positive modes $n>0$, as these
lower the energy (or momentum), which is not possible for the
vacuum. It is natural to expect that the unique ground state of
lowest $M$, the existence of which we assumed, is this vacuum
(they correspond in numerous examples). It certainly has the
lowest energy in this sector. Then our precise definition of
$\oint dx\,T$ in the vacuum sector is $\omega\Delta M=\oint
dx\,T/(2\pi)+c/24=L_0/R$ in the low energy region. However, at
present we do not have a general argument that the ground state is
invariant under $L_{-1}$, though $L_0|0\rangle=0$ holds for the
ground state by the preceding definition. We should note that,
although we can obtain the exact zero-energy many-particle
wavefunctions for our special Hamiltonian $H_0$, calculating the
norms of these, or operator matrix elements between them, remains
very difficult. There are also other ``sectors'' (subspaces) of
states in which one or more quasiholes is placed at the center of
the drop, and it seems in examples that all sectors that should
occur in the expected CFT arise in this way \cite{milr}. The
lowest angular momentum state in each sector must be annihilated
by all $L_n$ with $n>0$. The distinguishing feature of the vacuum
is that it is annihilated by $L_0$ and $L_{-1}$ also. The lowest
angular momentum state in each sector differs in $M$ from $M_0$ by
amounts of order $N$, and currently we do not have a way to
extract the expected finite difference of order 1 in conformal
weight (energy times $R$). It seems natural that the sectors with
higher $M$ for the same $N$ values are excited to positive
energies in the edge theory (suitably defined), but this has not
been proved. If we could confirm this, it would confirm the
identification of the ground state with the vacuum, because in a
unitary CFT the conformal weights of all states other than the
vacuum are positive.

The presence of other conserved quantities in the original LLL
model leads to further conserved quantities in the effective
theory of the edge. Thus, particle number is always conserved, and
there is a corresponding divergenceless Noether current in the
full theory of bulk and edge (though the expression for this LLL
current is not entirely obvious). Then there is also a
corresponding divergenceless current $j_z$ in the edge theory,
which is now chiral (its counterpart is $j_{\overline z}=0$) and
obeys $\overline\partial j_z=0$. This is present in all QH
systems. To make the discussion somewhat more general, let us
consider also the case in which the particles carry spin 1/2, the
special Hamiltonian is invariant under SU(2) acting internally on
the spin, and the ground state is an SU(2) singlet. (This serves
as a paradigm for larger Lie algebras acting as internal
symmetries, also.) Then there are local self-adjoint chiral spin
current operators $j_{az}$, where $a=1$, $2$, $3$ labels a basis
for the adjoint representation of SU(2). All such currents are
divergenceless, for example $\overline\partial j_{az}=0$, and
consequently have scaling dimension (conformal weight) 1
\cite{kz,aff85}. Because of conformal invariance, the most general
allowed form \cite{aff85} of the commutators (or the operator
products) of the currents are those for an affine Lie (Kac-Moody)
algebra for SU(2) \cite{kz}. Thus the currents are good conformal
fields (no logarithms in their correlators). The existence of such
currents in a unitary CFT implies that $c\geq 1$, ruling out the
trivial case $c=0$ among others. The affine Lie algebra contains a
single parameter $k$ which must be a positive integer (or $k=0$ in
the trivial case), as a consequence of the unitarity of the
theory. For the particle number or U(1) symmetry, the affine Lie
algebra simplifies as the structure constants vanish, and the
corresponding $k$ is positive but need not be an integer; instead,
given a suitable normalization of the U(1) current $j_z$, it
corresponds to the filling factor (Hall conductivity) $\nu$ of the
QH system. In the spin case, $k$ is a quantized Hall conductivity
for spin. Another way to state the affine Lie algebra symmetry,
similar to what we did for conformal invariance, is that if
$w^a(z)$ is a vector field transforming in the (co-)adjoint
representation, with $\overline\partial w^a=0$, then
$j_{wz}(z)=\sum_a w^a(z) j_{az}(z)$ is a divergenceless vector
field: $\overline\partial j_{wz}=0$.

\section{Example}

Next we will consider an example that at one time was considered
to be a counterexample to the notion (proposed in MR \cite{mr})
that the effective theory of the edge is a CFT and is the same one
that appears in describing the ground state wavefunction as a
correlator in a Euclidean CFT. This example is the Haldane-Rezayi
(HR) state \cite{hr}. It is the spin-singlet zero-energy ground
state of a special ``hollow-core'' LLL Hamiltonian for fermions
with spin 1/2. The ground state wavefunction is a Laughlin factor
times a correlator in the non-unitary CFT of symplectic fermions
\cite{mr}. Later, the complete space of zero-energy wavefunctions
was obtained \cite{milr}. The edge excitations above the ground
state were naturally understood in terms of the charge degree of
freedom associated with the current $j_z$, and neutral fermions
carrying spin 1/2 under the internal SU(2), and with conformal
weight one. The latter would be natural if the edge theory were
the non-unitary symplectic fermion theory. In order to construct a
unitary theory, a theory of spin 1/2 fermions with local spin
currents, but without Lorentz invariance was proposed \cite{milr};
the correlators of the spin currents contained logarithms and did
not have the affine Lie algebra form. An alternative proposal
\cite{gfn} for the fermion sector of the theory was a unitary
Dirac fermion CFT, which gives the same Hilbert space and energy
spectrum up to a shift in the zero of energy, but has nonlocal
SU(2) spin currents \cite{sathya} (the conformal weight of the
fermions is then 1/2, because the vacuum is identified with a
distinct sector from the ground state above; this construction is
possible only because {\em two} sectors possess ground states
annihilated by the respective $L_{-1}$ in each fermion theory,
when zero modes are absent as they are here). The upshot of all
these works seems to be that no fully local unitary theory for the
edge of the HR state exists, and this is in part a consequence of
the (conformal) spin-statistics theorem in $1+1$ dimensions for
unitary Lorentz invariant fermion theories. The apparent conflict
with the arguments of the present paper is resolved because of the
argument in Ref.\ \cite{rg} that the bulk of the system is not
fully gapped; the fermion sector is gapless. Then it is not clear
that there is any gapless edge spectrum (described perhaps by the
zero-energy eigenstates of $H_0$) distinct from the general
gapless excitations of the bulk. Several other cases of
non-unitary CFTs (as well as some with other pathologies) that
arise in a similar way in QH systems are also argued to be gapless
in the bulk \cite{rr1,rg}. No examples are known to me in which a
non-unitary CFT arises from the bulk wavefunction, and the bulk is
known to be fully gapped (but note that in general deciding on the
latter issue may not be possible in numerical work because of
finite size restrictions, and it requires the construction of a
theory for the bulk, which for many QH systems has not been done).

\section{Further discussion}

Now we return to the general discussion. If we consider all the
local operators in the edge theory, then it makes sense to
classify them according to representations of the conformal
(Virasoro) algebra. Like the states of the edge theory, they form
highest-weight representations, which means that their conformal
weight ($L_0$ eigenvalue) cannot be lowered indefinitely.
Moreover, by applying a local field to the vacuum state, one
obtains another state. The two-point correlation function of pairs
of local operators then becomes essentially the inner product of
the corresponding states, and is finite (possibly zero). Hence the
local operators correspond to states in a positive definite space,
on which the Virasoro algebra acts \cite{bpz,cft}. It follows that
they must form unitary representations of the algebra, as the
states do. In particular, the representation generated by a
highest weight (such an operator is called a conformal primary
field) must be irreducible. This eliminates logarithmic effects
from all correlators of such fields. Further, in a unitary CFT,
the conformal weight of a primary field must be non-negative (it
can be zero only for the identity operator).

In our present situation, all local operators $\cal O$ obey
$\overline\partial {\cal O}=0$, and there are corresponding
conserved quantities. These operators acting on the vacuum span
the vacuum sector of states, and there is an operator
corresponding to each state. These operators form an algebra
called the chiral algebra of the QH edge system. The chiral
algebra contains the Virasoro and U(1) particle number current
algebras, the current algebra for spin and so forth (if any), and
also further operators. (We should note that the remaining sectors
of states other than the vacuum sector are not created from the
vacuum by using local operators. Indeed, in the QH system, they
must be created by acting with some quasihole operator located
somewhere in the bulk. They can be viewed as corresponding to
states in the non-vacuum sector, nonetheless.)

As examples of local primary operators other than $T$ and the
currents, we can consider the particle creation and annihilation
operators. These are local as the corresponding operators (which
are projected to the LLL) are local in the full bulk theory, and
can be applied close to the edge. If the underlying particles are
bosons (a situation that may possibly be realizable in rotating
cold trapped atoms), then these operators commute when well
separated, and similarly for fermions (as in QH systems), they
anticommute. This can be extracted from correlation functions
(defined using time-ordered products, as usual), which must be
single valued in both cases, which leads to (anti-)commutation of
the operators, and hence using conformal invariance the conformal
weight of the primary field must be an integer in the boson case.
Indeed, this is true for all strictly local operators. (The
single-valuedness follows by considering the function first in the
full bulk theory, where it follows from the general definition of
the particle field operators, say using imaginary time; it is not
lost by passing to large separation and low energies, near the
edge, to obtain the corresponding function in the edge theory.
This argument holds for any effective theory of the edge, even
without conformal invariance, not only for the special Hamiltonian
used here. Hence the claim in Ref.\ \cite{zpm} that the operators
for electrons in a chiral effective edge theory might not
anticommute at large separation and equal time is incorrect.) In
the case when the underlying particles are fermions, the field
operators in the edge theory obey anticommutation relations, and
the correlation functions are single-valued, but the fermion field
operators must have half-odd-integer conformal weight. (Sometimes
fermion fields are viewed as not strictly local, because they do
not commute at large spatial separations.) Then the natural
boundary condition on these fields, which must be the same in all
sectors, in the cylindrical spacetime we are considering, is
antiperiodic \cite{cft}. In this situation the chiral algebra in
fact becomes a chiral superalgebra (in the ``Neveu-Schartz''
sector), as some of the fields obey anticommutation relations.
Examples include the Laughlin states for fermions at $\nu=1$,
$1/3$, \ldots, and the Moore-Read state at $\nu=1/2$ (or $5/2$).
When the conformal weight of the particle field is $3/2$, it
generates a superconformal algebra (which also includes $T$ and
the current $j_z$) at least in several examples, including the
Laughlin $1/3$ state, the Moore-Read $1/2$ state, and all other
members of the Read-Rezayi series with $M=1$ also
\cite{mr,milr,rr2}.

\section{Beyond special Hamiltonians}

Now we turn briefly to the problems that occur beyond the realm of
special Hamiltonians. First, for more general local bulk
Hamiltonians that can be deformed to some special Hamiltonian
without crossing a bulk phase transition, the (fully-gapped)
topological phase in the bulk remains the same as for the special
Hamiltonian. In this situation, the edge theory should remain
chiral and local, but will in general be perturbed by addition of
terms (integrals of a local operator) to the Hamiltonian as the
bulk Hamiltonian is perturbed away from the special one.
(Actually, changes in the Hamiltonian near the edge can cause edge
reconstruction, in which case the edge does not remain chiral; we
do not consider this here.) Then the Hamiltonian density of the
edge theory may contain extra local operators added to the stress
tensor $T$ of the CFT. Such perturbations may be classified as
relevant, irrelevant, or marginal under the renormalization group.
These occur when the scaling dimension (which is the same as the
conformal weight in the present chiral case) is $>2$, $<2$, or
$=2$, respectively. Irrelevant operators produce no essential
change and can be dropped at sufficiently low energies. Relevant
ones cause a flow to some other fixed point, which is still a
chiral theory. We note that in a chiral theory, the perturbations
generally break Lorentz invariance. The low-energy fixed point may
still be conformal, perhaps with some redefinitions (for example,
an interesting example is two chiral Majorana fermi fields
$\psi_1$, $\psi_2$, perturbed by a coupling $im\psi_1\psi_2$ where
$m$ has dimension $1$; the low-energy fixed point is conformal
provided one shifts the zero of momentum for the fermions by
constants proportional to $\pm m$).

Marginal perturbations can be classified as marginally relevant,
marginally irrelevant, and exactly marginal. The first two have
similar effects as the relevant and irrelevant cases,
respectively. In the exactly marginal case, the coefficients of
the perturbation are not renormalized at any order in perturbation
theory. As examples, any chiral CFT has an exactly marginal
perturbation of dimension $2$, namely that by the stress tensor
$T$. This has the effect of changing the velocity of all the edge
excitations away from $v$, thus violating the Lorentz invariance
of the original theory, though of course it is replaced by another
Lorentz invariance. If the CFT is a tensor product of theories,
which means that the stress tensor is a sum of those in two
decoupled theories, $T=T_1+T_2$ (the theories may still be coupled
through superselection rules on the non-vacuum sectors or
quasiparticle operators), then adding a multiple of say $T_1$ to
the Hamiltonian $T$ changes the velocity in theory $1$ but not in
theory $2$. More generally, such a theory can be perturbed by an
operator ${\cal O}_1{\cal O}_2$ with a factor ${\cal O}_i$, $i=1$,
$2$ from each theory, with ${\rm dim}\,{\cal O}_1+{\rm dim}\,{\cal
O}_2=2$ to ensure it is marginal. If it is exactly marginal, this
can have an effect similar to changing the velocities of the two
theories, but may also cause mixing of the modes. For example,
with more than one chiral boson field $\phi_1$, $\phi_2$, \ldots,
as in various abelian QH states, a term
$\partial\phi_1\partial\phi_2$ can be added, and is exactly
marginal (see e.g.\ Ref.\ \cite{hald95}). Such problems must be
treated case by case, and we cannot go further here. But the point
is that for such a Hamiltonian in the bulk, the edge theory is not
strictly conformal, but is a perturbation of a CFT.

If there is no known special Hamiltonian that produces a
particular topological phase, such as in many cases including
Abelian states believed to occur in the QH system, the edge theory
may or not be chiral (it must still be local and unitary when the
bulk Hamiltonian is). For chiral cases, the preceding discussion
of perturbations of a CFT still applies. We have no general
argument that the edge theory for a generic Hamiltonian in such a
phase must be a perturbation of a chiral CFT, but we expect this to
be the case. However, if one can tune parameters of the theory so
that all low-energy edge excitations propagate with a common
velocity, while remaining in the same bulk topological phase, then
the arguments of this paper still go through. For non-chiral
cases, the discussion of perturbations of a CFT is similar, though
now operators perturbing a CFT may contain parts from both right-
and left-moving sectors, but must still be local. Again, see Ref.\
\cite{hald95} for exactly-marginal examples. In these cases, it is
less clear if there is a way to establish the existence of an
underlying CFT. Finally, in all cases one can also consider
long-range perturbations of the bulk theory, such as the Coulomb
interaction, which leads to a long-range current-current
interaction in the edge theory, which spoils Lorentz invariance.

\section{Some recently-proposed trial states}

Now we turn to an application of the above argument, that the effective field theory of the edge of a gapped topological phase must be both conformally invariant and unitary, to some trial states of recent interest. We will find that for infinitely many of these the number of edge excitations they would have is not compatible with any unitary theory.

The trial wavefunctions in question \cite{srcb,bernhald1,bernhald2,bernhald3} are generalizations of the paired and clustered states of spin-polarized particles, such as those in Refs.\ \cite{laugh,mr,rr2}. In the simplest cases, which always describe bosons, the functions are defined by requiring them to vanish as the $R$th power of separation ($R=2$, $3$, \ldots) as $k+1$ particles come to the same point ($k=1$, $2$, \ldots). For small enough $R$ or $k$, this produces a unique ground state, and reasonable numbers of excited edge (and also quasihole) states \cite{src}. For $k>1$ and $R>3$, further conditions are required, which may be described in terms of the relation with ``$(k,R)$ admissible partitions''; for $R>2$ it is required that $k+1$ and $R-1$ be coprime. (Further states are easily obtained from any of these by multiplying by a power of the Laughlin-Jastrow factor.) These functions are all given by Jack polynomials in the coordinates $z_i$ \cite{feigin,bernhald1}. Then the $R=2$ cases are the states labeled by $k$ (and $M=0$) in Ref.\ \cite{rr2}, where $k=1$ is the Laughlin state \cite{laugh}, and $k=2$ is the Moore-Read state \cite{mr}. The case $k=2$, $R=3$ is the so-called Gaffnian state \cite{srcb}. Apart from a power of the Laughlin-Jastrow factor, the latter is a conformal block (correlator) in the non-unitary CFT known as the $M(5,3)$ minimal model \cite{srcb} (we follow the conventions of Ref.\ \cite{cft}). It is known  \cite{feigin} that these Jack polynomials are related in this way to minimal models for $k=2$ (and all $R$), and to minimal models for $W_k$-algebras (the $W_2$ algebra is simply Virasoro) for $R=2$ (and all $k$); it is conjectured \cite{feigin} to hold for all $k$ and $R$.  [Including the charge sector, the fusion rules for the bulk are known to be those of SU$(R)$ level $k$ \cite{ardonne}.] This conjecture is supported by recent work \cite{bgs}. All cases with $R>2$ are non-unitary CFTs. In the cases at $k=1$ or $R\leq3$, there are (local) special Hamiltonians involving interactions among at most $k+1$ particles for which these states are (at least conjecturally) the unique zero-energy ground states. Bernevig and Haldane \cite{bernhald1} state that for $k>1$ and $R>3$ they did not find a local Hamiltonian with $n$-body interactions for any $n$ that produces the Jack trial wavefunctions and no others. Even when such a Hamiltonian exists, it is not established whether it is gapped or gapless in the thermodynamic limit. The authors of Ref.\ \cite{srcb} acknowledge that their special Hamiltonian may be gapless in this limit.

We will consider for the case $k=2$ the ``counting formulas'' that give the dimension of the spaces of trial wavefunctions for edge excitations of a disk as discussed above. Earlier results of this type, and also some for zero-energy quasihole states on a sphere, were obtained for $R=2$ and all $k$ (with no use of CFT) in Refs.\ \cite{milr,rr1,rr2,aks,read06,srcb}, prior to Refs.\ \cite{bernhald1,bernhald2,bernhald3} [also, the connection of the counting formulas with those for $(k,2)$ partitions was pointed out in Ref.\ \cite{read06}]. Simon {\it et al.}\ \cite{srcb} give a general formula for the number of zero-energy states (for their special Hamiltonian) for $n$ flux quanta added to the ground state on the sphere [see eq.\ (5) in Ref.\ \cite{srcb}]. Using techniques outlined in Ref.\ \cite{read06}, this can be generalized to count states at each angular momentum $L_z$, then used to obtain the number of edge states (if such they be), by taking the limit in which first $n\to\infty$, and then $N\to\infty$. The results can be presented as a generating or partition function, which is a series in an indeterminate $q$, in which the coefficient of the term $q^{\Delta M}$ is the dimension of the subspace of zero-energy states with angular momentum change $\Delta M$ from the ground state (as above). The result for the edge of the Gaffnian state is%
\be%
\frac{1}{(q)_\infty}\sum^\infty_{f=0}\frac{q^{f(f+1)}}{(q)_{2f}},
\label{m53char}\ee%
where $(q)_m=(1-q)(1-q^2)\cdots(1-q^m)$. We note that $1/(q)_\infty$ is the partition function for the edge of the Laughlin state \cite{milr}, and is always present in such formulas because of the ubiquity of the density excitations at the edge. The remaining factor can be identified as the character $\chi_{1,1}^{5,3}$ for the vacuum (or identity) sector of the $M(5,3)$ minimal model [see eq.\ (4.4) in Ref.\ \cite{kkmm}]. It seems to be a general result that the counting of edge states in the thermodynamic limit agrees with the vacuum sector of the CFT that underlies the trial wavefunctions.

We now recall more details about the minimal CFTs for the Virasoro algebra \cite{bpz}, which in
general are denoted $W_2(p,p')=M(p,p')$, parametrized by a pair of coprime
positive integers $p$, $p'$ with $p>p'$ \cite{cft}.
For the minimal models, the primary fields $\phi_{r,s}$ are labeled by ordered pairs of integers $(r,s)$, with $1\leq r\leq p'-1$, $1\leq s \leq p-1$ (the so-called minimal block in the Kac table). There is an equivalence $r\to p'-r$, $s\to p-s$, and each inequivalent primary occurs just once in the theory. The conformal weights and central
charge are given by \cite{bpz}%
\bea%
c&=&1-6\frac{(p-p')^2}{pp'},\\
h_{r,s}&=&\frac{(pr-p's)^2-(p-p')^2}{4pp'}.\eea%
The unitary minimal models, which are the only unitary CFTs with $c<1$, are those in which $p'=m$, $p=m+1$, with $m=2$, $3$, $4$, \ldots (the case $m=2$ is trivial). In these $c=1-6/[m(m+1)]$.

The effective central charge for a CFT, or for anything with similar counting formulas, can be defined in terms of the rate of growth of the number of states at high $\Delta M$ in the vacuum (or any other) sector. In a modular-invariant CFT, such as any of the $W_k$ minimal models, it can be obtained by using a modular transformation \cite{cft}, and is given by $c_{\rm eff}=c-24h_{\rm min}$, where $h_{\rm min}$ is the lowest conformal weight in the theory. In unitary theories, $h_{\rm min}$ is zero, so $c_{\rm eff}=c$. For the Virasoro minimal models, $h_{\rm min}$ is the lowest value of $h_{r,s}$ in the minimal block. Using Bezout's lemma, and the fact that $p$ and $p'$ are coprime, one obtains $h_{\rm min}=[1-(p-p')^2]/(4pp')$, which is negative in all non-unitary cases. That is,%
\be%
c_{\rm eff}=c-24h_{\rm min}=1-\frac{6}{pp'}.\ee%
For the counting of QH trial wavefunctions, we will always remove the contribution $1$ to both $c$ and $c_{\rm eff}$ that is due to the charge sector. Then for the Gaffnian case $M(5,3)$ the central charge is $c=-3/5$, while the effective central charge is $c_{\rm eff}=3/5$.
More generally, the series of Jack states with $k=2$ are known to be (apart from the charge sector) conformal blocks of the minimal model $M(R+2,3)$ \cite{feigin}, and the counting formulas for these lead to the effective central charge $R/(R+2)$  \cite{fjm}, as we discuss further below. [A numerical check was carried out in Ref.\ \cite{bernhald3}; note that these authors call this quantity the ``thermal Hall conductivity'', on the assumption that the bulk of the system is gapped (for some Hamiltonian) \cite{rg}.]

Now we come to the crucial point. Even if the trial wavefunctions are conformal blocks in a non-unitary CFT, they might describe a gapped bulk phase and its edge excitations. If there is a special Hamiltonian
for which the trial functions are zero-energy eigenstates, and which has a gap in the bulk, then we have seen that the effective theory of the edge must be a unitary CFT, which must be different from that used in the trial wavefunctions if the latter is non-unitary. In this case, its $c_{\rm eff}$ value, defined by counting the excitations, must equal the central charge $c$ of the unitary CFT. For the cases with $k=2$, this value is $c=R/(R+2)$. As this is less than one, the allowed values of $c$ for a unitary theory are very restricted, and one finds that most values of $R$ do not correspond to such a value. The only values of $R$ for which $c=R/(R+2)$ corresponds to a unitary representation of Virasoro are given by%
\be%
R+2=m(m+1)/3\ee%
with either $m$ or $m+1$ divisible by $3$ (but not by $9$).
After the Moore-Read case $R=2$, the next such values are $R=8$, $12$, $42$, \ldots. That is, for $R\neq m(m+1)/3-2$ there is {\em no unitary CFT that is an acceptable candidate} to be the edge theory of these trial functions. In particular, this argument applies to the Gaffnian [$M(5,3)$ minimal model], for which a special Hamiltonian does exist. We conclude that it cannot describe a gapped topological phase,
a conclusion also discussed in Refs.\ \cite{srcb,read08b}. Most likely, the special Hamiltonian is gapless in the thermodynamic limit.

For the cases $R= m(m+1)/3-2$ that are not excluded by the above argument, we can examine the counting of states in more detail, if we can show that the generating or partition function of the edge excitations is given by the character for the $M(R+2,3)$ minimal model. For the primary field labeled $r$, $s$, this character can be written in the form \cite{cft}%
\be%
\chi_{r,s}^{p,p'}=\frac{1}{(q)_\infty}\sum_{\lambda=-\infty}^\infty
\left(q^{\lambda^2pp'+\lambda(pr-p's)}-q^{(\lambda p'+r)(\lambda p+s)}\right)\label{char}\ee%
(we omit the factor $q^{h_{r,s}}$ that is usually included). In every case, it can also be written in a ``fermionic'' form  \cite{welsh}, similar to but usually more complicated than that for the $M(5,3)$ example in eq.\ (\ref{m53char}) above. Our interest is in $r=s=1$.
For the edge excitations of the Jack states, or the $(2,R)$ admissible partitions, the information given in Ref.\ \cite{bernhald3} for $k=2$ is insufficient to answer the question, and contains undefined notation. Fortunately, the full results and notation were given by Feigin and co-workers \cite{fjm}, and using their results one can show that as $N\to \infty$ (with $N$ even) the counting of the $(2,R)$ admissible partitions and of the corresponding polynomials exactly coincides with the fermionic form of the character $\chi_{1,1}^{R+2,3}(q)$ for all $R$, after removing the factor $1/(q)_\infty$ for the charge sector (the value of $c_{\rm eff}$ then follows). The remaining question at issue is whether these characters can be equal for distinct pairs $p$, $p'$ of coprime integers ($p>p'>0$) such that $pp'$ is the same (in our particular case, one of the pairs is unitary, $p=m+1$, $p'=m$). However, one can see that given a series of the form (\ref{char}), the values of $p$, $p'$ are uniquely determined, which rules out distinct pairs. Hence for $k=2$ no unitary CFT exists for {\em any} value $R>2$.

For the trial functions (Jack polynomials) with $k=2$, $R>3$, there appear to be no local special Hamiltonians. The argument still shows that there can be no unitary CFT that accounts for the ``edge'' trial states when $R>2$. In the absence of a Hamiltonian that is local and produces these zero-energy states and no others, it is unclear if it is meaningful even to ask if these functions represent edge excitations of a topological phase. The argument would still apply for a Hamiltonian that is gapped in the bulk, but for which the trial states are not eigenstates, if it is claimed to be in the same phase as those trial states (and even when it has different velocities for the charge and other excitations on the edge). If there is some non-local Hamiltonian that has a bulk gap and these trial states as zero-energy edge states, then the edge effective field theory still cannot be conformally invariant, which would not be surprising in view of the non-locality. Such a non-local theory would presumably be physically unacceptable in any case.

A last possibility might be that the trial ground state for the disk corresponds to an excited state on the edge, and vice versa, as mentioned for the Haldane-Rezayi state earlier. This seems to be inconsistent in all cases, as the identity in the fusion rules for bulk excitations is fixed. The fusion rules \cite{ardonne} and the identification of the identity for the bulk and the edge must agree.

For the models with $k>2$, the argument of this section does not go through as the values of $c_{\rm eff}$ are bigger than $1$ (for $R>2$), where the constraint of unitarity on Virasoro representations is very weak. It would be natural to try to generalize it using analogous statements for $W_k$ minimal models, if we could show directly that the necessary local conserved higher spin currents (of weights $3$, $4$, \ldots, $k$) exist in the quantum-mechanical edge theories of these states. The unitary $W_k$ minimal models have central charge of the form%
\be%
c=(k-1)\left(1-\frac{k(k+1)}{m(m+1)}\right),\ee%
where $m=k+1$, $k+2$, \ldots \cite{fatluk}. In the lowest case $m=k+1$, the $W_k$ minimal model is the ${\bf Z}_k$ parafermion theory used in Ref.\ \cite{rr2}. Then for the $R=2$ cases, which are the Read-Rezayi states \cite{rr2}, the unitary boundary theory does exist, and including the charge sector is SU$(2)$ level $k$; the higher spin currents must exist in the low-energy effective theories of these edges, which lends some support to the idea.

Bernevig and Haldane conjecture \cite{bernhald3} that the effective central charge values arising from the Jack polynomial states are given by $c_{\rm eff}=(k-1)R/(R+k)$ in all cases. These agree with $c$ in a unitary $W_k$ minimal model only if%
\be%
R+k=\frac{m(m+1)}{k+1}.\ee%
Clearly, most values of $R=3$, $4$, \ldots (with $R+k$ and $k+1$ coprime) do not satisfy this condition. However, at present we do not know that the $W_k$ higher-spin currents necessarily exist, even when a special Hamiltonian does.  Hence we have not yet eliminated the possibility that some other unitary CFT exists that would agree with the values of $c_{\rm eff}$ in the cases $k>2$, $R>2$.

\section{Conclusion}

To summarize, a chiral edge theory of a local Hamiltonian that has a gapped spectrum in the bulk (in the thermodynamic limit) and a single velocity for the edge excitations must be both conformally invariant and unitary. This applies in particular for many known special Hamiltonians for which the edge excitations can be identified as zero-energy eigenstates whose angular momentum differs from that of the ground state by amounts of order $1$. Then for many examples of proposed trial wavefunctions, including the Gaffnian state for which a local special Hamiltonian is known to exist (but is not known to be gapped in the bulk in the thermodynamic limit), the possibility that they represent a topological phase is ruled out because the number of edge excitations is not compatible with a unitary conformal field theory. The only remaining possibility for an interpretation of these states, if some Hamiltonian exists for which they are either exact eigenstates, or are in the same phase, is that either some excited states approach zero energy in the thermodynamic limit, or the Hamiltonian is significantly non-local.

We are grateful to N. Cooper and S. Simon for helpful discussions and correspondence. This work was supported by NSF grant no.\ DMR-0706195.

\end{document}